\documentclass[pre,twocolumn,a4paper,superscriptaddress,floatfix]{revtex4}
\usepackage{epsfig}
\usepackage{amssymb}
\begin{document}

%
%
\title{Phase transition and landscape statistics of the number
       partitioning problem}
\author{Peter F. Stadler}
\affiliation{Bioinformatik, Institut f{\"u}r Informatik.
            Universit{\"a}t Leipzig, Kreuzstra{\ss}e 7b,
            D-04103 Leipzig, Germany}
\affiliation{Institut f{\"u}r Theoretische Chemie und Molekulare
            Strukturbiologie, Universit{\"a}t Wien,
            W{\"a}hringerstra{\ss}e 17, A-1090 Wien, Austria}
\affiliation{The Santa Fe Institute, 
            1399 Hyde Park Road, Santa Fe, NM 87501, USA}
\author{Wim Hordijk}
\affiliation{Instituto de F\'{\i}sica de S\~ao Carlos, Universidade de S\~ao
Paulo, Caixa Postal 369, 13560-970 S\~ao Carlos SP, Brazil.}  
\author{Jos\'e F. Fontanari}
\affiliation{Instituto de F\'{\i}sica de S\~ao Carlos, Universidade de S\~ao
Paulo, Caixa Postal 369, 13560-970 S\~ao Carlos SP, Brazil.}  

\begin{abstract}
The phase transition in the number partitioning problem (NPP), i.e., the
transition from a region in the space of control parameters in which almost
all instances have many solutions to a region in which almost all instances
have no solution, is investigated by examining the energy landscape of this
classic optimization problem. This is achieved by coding the information
about the minimum energy paths connecting pairs of minima into a tree
structure, termed a barrier tree, the leaves and internal nodes of which
represent, respectively, the minima and the lowest energy saddles
connecting those minima. Here we apply several measures of shape (balance
and symmetry) as well as of branch lengths (barrier heights) to the barrier
trees that result from the landscape of the NPP, aiming at identifying
traces of the easy/hard transition. We find that it is not possible to tell
the easy regime from the hard one by visual inspection of the trees or by
measuring the barrier heights. Only the {\it difficulty} measure, given by
the maximum value of the ratio between the barrier height and the energy
surplus of local minima, succeeded in detecting traces of the phase
transition in the tree.  In adddition, we show that the barrier trees
associated with the NPP are very similar to random trees, contrasting
dramatically with trees associated with the $p$ spin-glass and random
energy models.  We also examine critically a recent conjecture on the
equivalence between the NPP and a truncated random energy model.

\end{abstract}

\maketitle

\section{Introduction} \label{sec:Introduction}

The relevance of the concepts and techniques of statistical physics to
understanding the typical behavior of classes of optimization or decision
problems had been pointed out by many authors already in the middle of the
1980s (see, e.g., Refs.  \cite{Anderson:86,Mezard:87}). However, it was
only about ten years later, owing mainly to the finding of a ubiquitous
peak in computational cost signaling a transition between easy and
difficult instances of optimization problems, that the physics approach has
succeeded to attract the attention of the computer science community (see,
e.g., Refs. \cite{Kirkpatrick:94,Hogg:96,Hayes:97,Monasson:99,Martin:01}).
In particular, instances in the phase transition region are now routinely
used to benchmark algorithms and search heuristics, and so a precise
location of the critical point in addition to the estimate of the width of
the transition region has gained considerable practical importance.

The specific optimization problem we consider here is the number
partitioning problem (NPP), which is one of the basic NP-complete
problems that form the core of the theory of NP-completeness
\cite{Garey:79}. NPP has an easy formulation: Given $N$ not necessarily
distinct positive numbers $a_1$, \dots, $a_N$ find a subset
$\mathfrak{X}\subset\{1,\dots,N\}$ such that
\begin{equation}
E(\mathfrak{X}) = \left| \sum_{j\in\mathfrak{X}} a_j -  
                         \sum_{j\notin\mathfrak{X}} a_j \right|
\end{equation}
is minimized. We remark that NPP can be regarded as a Mattis-like
Ising spin model with Hamiltonian
\begin{equation}\label{eq_Mattis}
H(\mathfrak{X})=E^2(\mathfrak{X}) = \sum_{i,j} a_ia_j \sigma_i\sigma_j
\end{equation}
where $\sigma_i=+1$ if $i\in\mathfrak{X}$ and $\sigma_i=-1$ if
$i\notin\mathfrak{X}$ \cite{Fu:89}. It is therefore natural to consider $E$
as an energy (cost) landscape over the hypercube; in other words, single
spin flips are a natural way of defining a neighborhood relation for the
NPP.  For concreteness, we will assume from here on that the $a_i$s are
independent, identically distributed random variables that take on integer
values between $1$ and $l$ with equal probability.

A partition $\mathfrak{P}$ is \emph{perfect} if $E(\mathfrak{P})=0$ or $1$
for $\sum_i a_i$ even or odd, respectively. The existence of perfect
partitions depends on the accuracy to which the numbers $a_i$ are
determined as well as on the size of the problem $N$. The crucial control
parameter here is the ratio between the number of bits to which $a_j$ is
specified and the problem size
\begin{equation}
\kappa = \frac{\log_2 l}{N}.
\end{equation}
The relevance of $\kappa$ can be appreciated by considering the annealed
estimate of the expected number of perfect partitions, ${\mathcal S} =
2^N/l$, so that $\kappa = 1 - (\log_2 {\mathcal S})/N$
\cite{Gent:98}. Hence the annealed theory indicates that for $\kappa<1$
there is an asymptotically exponential number of perfect partitions, while
for $\kappa>1$ the probability of finding a perfect partition is
exponentially small. Extensive numerical simulations and statistical
mechanics calculations corroborate the value $\kappa_c = 1$ as the
threshold separating the easy-to-solve from the hard-to-solve regimes
\cite{Gent:98,Mertens:98}.

Probabilistic and statistical mechanics analyses of the ground states of
the Hamiltonian (\ref{eq_Mattis}) in the limit of infinite precision $l \to
\infty$, in which the $a_i$s can be viewed as continuous variables, have of
course failed to detect the phase transition (see, e.g., Refs. \cite{Fu:89,
Karmarkar:86,Ferreira:98}).  The thermodynamics for general $l$ was
`solved' under the assumption of self-averaging of the random variables
$a_i$, but that solution is not completely satisfactory since it predicts a
negative entropy for $\kappa > \kappa_c$ \cite{Mertens:98}.  Thus a
reliable theory for the ground states of the NPP, probably based on the
replica method, has yet to be obtained.

Generally, the aforementioned phase transition is defined as a transition
from a region in the space of control parameters in which almost all
instances have many solutions to a region in which almost all instances
have no solution \cite{Kirkpatrick:94,Hogg:96}. Hence, the investigations
have relied mainly on exhaustive search procedures, such as
branch-and-bound algorithms, that guarantee the finding of the global
optima, or on statistical mechanics calculations of the expected
properties, such as the entropy, of the ground states.  In this
contribution we seek evidence of this easy/hard transition in the structure
of the cost landscape of the optimization problem, focusing on the
distribution of optima and on the distribution of cost barriers between
these optima. To this end, we code the information about the paths of
minimal cost leading to different optima in a tree structure, termed the
{\it barrier tree} of the cost landscape. The leaves of this tree represent
the local optima and the internal nodes the lowest-cost saddles connecting
those optima.  Barrier trees have been widely used to study protein
\cite{Becker:97,Wales:98,Garstecki:99}, RNA \cite{Flamm:00a,Flamm:02a}, and
spin-glass \cite{Nemoto:88,Ferreira:00a,Fontanari:02a,Hordijk:02}
landscapes.

We find that the structure of the landscape, as measured by the local
minima and their connecting saddle points, shows surprisingly little
difference in the easy and hard regimes.  The sharp transition between
these two regimes is revealed only by the \emph{difficulty} of the
landscape, a parameter measuring the maximum ratio of energy barrier to
energy gain for the escape from a metastable state, which is directly
related to the optimal speed of convergence of simulated annealing
\cite{Hajek:88,Catoni:92,Kern:93,Ryan:95,Catoni:99}.  We stress that our
goal here is not to locate the transition point, which can be achieved by
simply looking at the value of the global energy minimum in an ensemble of
randomly generated instances (in this sense, a barrier tree always contains
the information on whether the given instance is easy or hard), but to seek
evidences of the easy and hard regime on other global statistical
properties of the energy landscape.  Furthermore, we examine a remarkable
conjecture about the equivalence between the infinite accuracy version of
the NPP and the random cost problem \cite{Mertens:00}, and show that,
despite the equivalence at the level of the energy distributions, their
barrier trees are completely different.

\section{Barrier Trees}

\begin{figure}
  \begin{center}
    \begin{tabular}{cc}
      \includegraphics[width=0.24\textwidth,clip=]{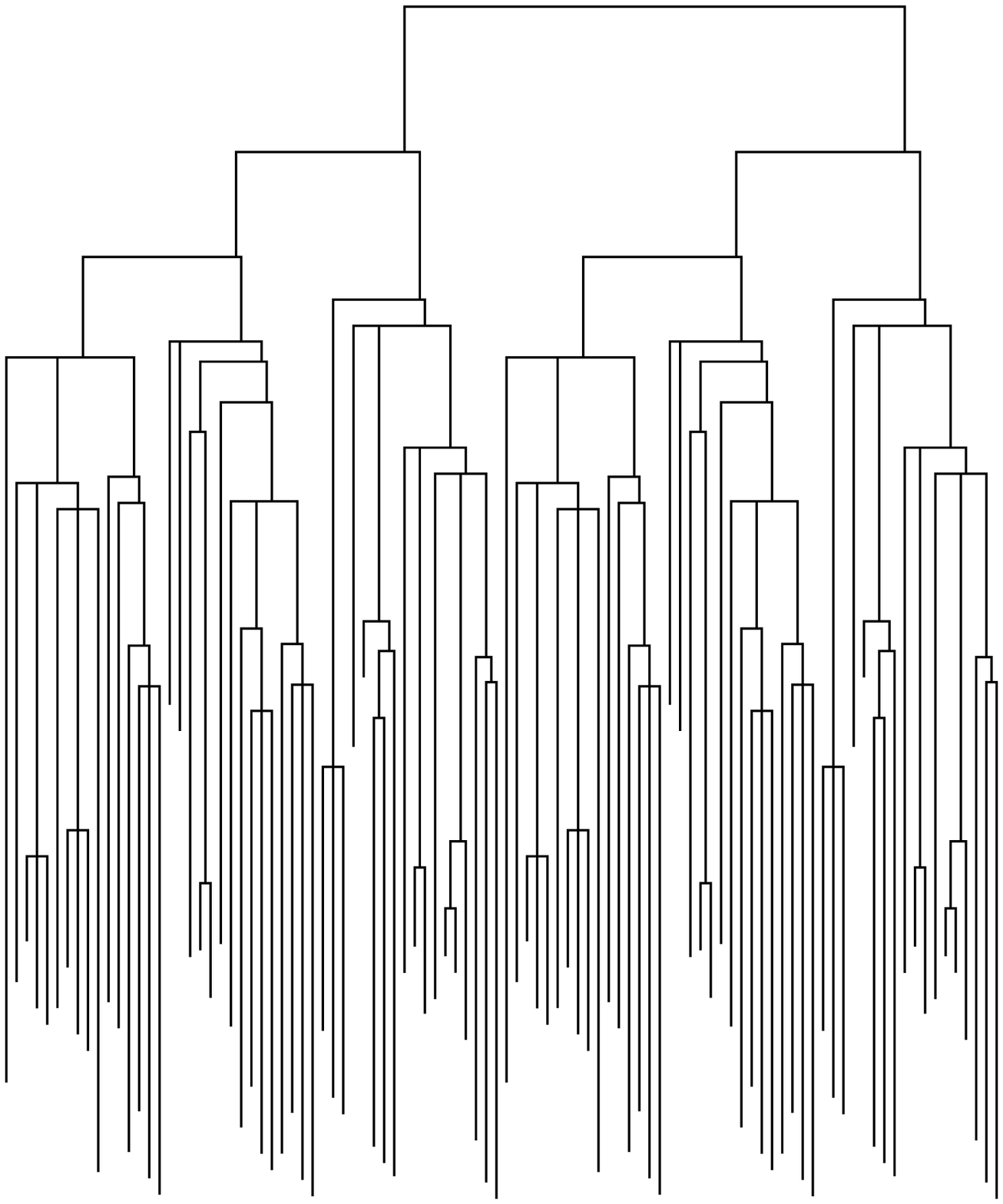} & 
      \includegraphics[width=0.24\textwidth,clip=]{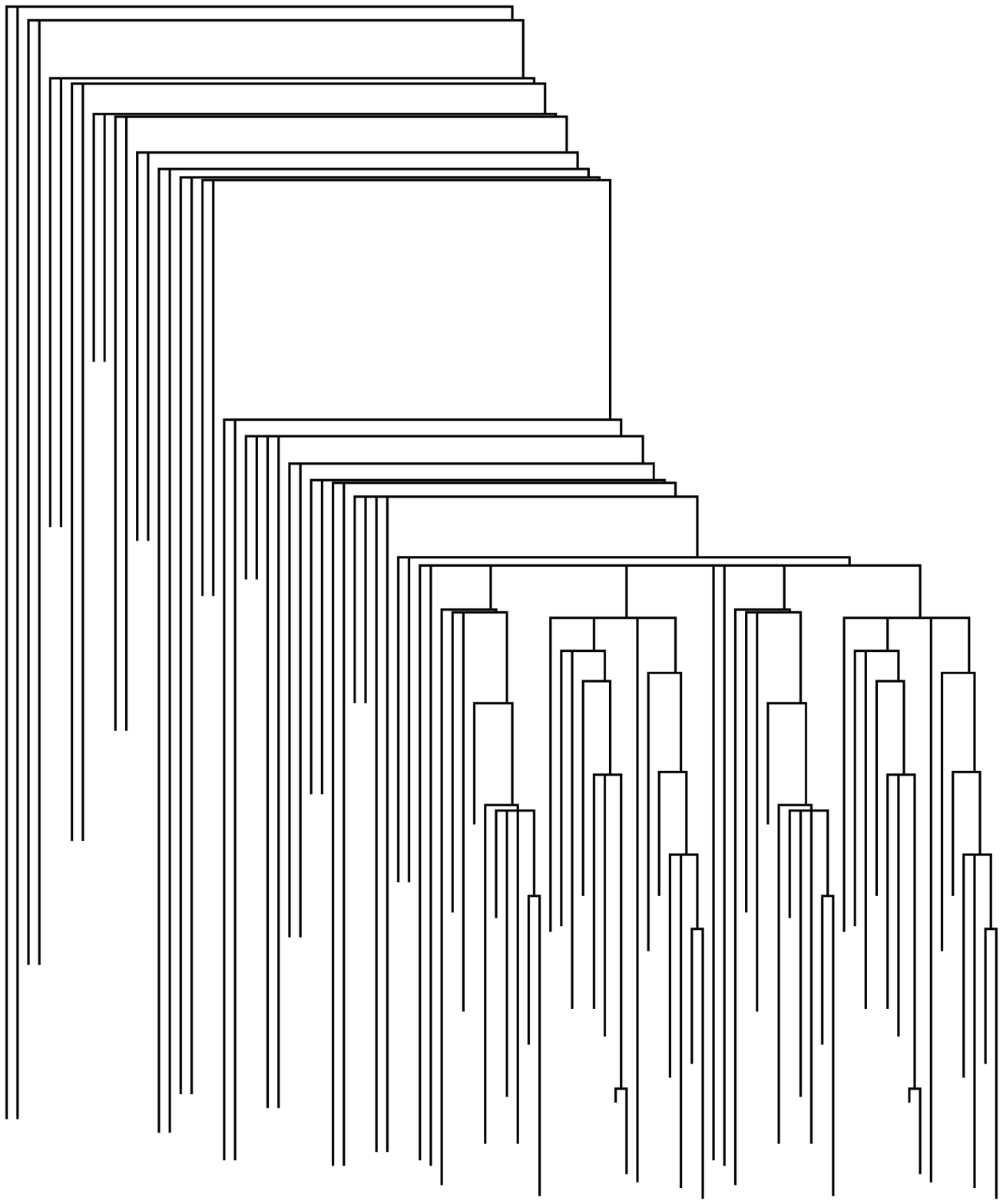} \\
      {\small(a)} & {\small(b)} 
    \end{tabular}
  \end{center}
  \caption{Typical barrier trees for
(a) the number partitioning problem with infinite precision numbers and 
(b) the truncated random energy model (random cost problem)
for $N=10$.}
  \label{fig:trr_diff}
\end{figure}

The energy of the lowest saddle point separating two local minima $x$
and $y$ is
\begin{equation}
  E[x,y] = \min_{ \mathbf{p}\in\mathbb{P}_{xy} } \,
           \max_{ z \in \mathbf{p} } E(z)
\label{eq:saddle}
\end{equation}
where $\mathbb{P}_{xy}$ is the set of all paths $\mathbf{p}$ connecting $x$
and $y$ by a series of subsequent spin-flips \cite{Nemoto:88,Vertechi:89}.
The \emph{barrier height} $B(x)$ of a metastable state $x$ is the minimum
height of a saddle point that connects it with metastable state $y$ with
strictly smaller energy $E(y)<E(x)$. In symbols,
\begin{equation}
  B(x) = \min\left\{ E[x,y] - E(x) \big| ~y : E(y)<E(x) \right\} .
\label{eq:defdepthC}
\end{equation}
Since a direct evaluation of eq.(\ref{eq:saddle}) would require the
explicit constructions of all possible paths it does not provide a feasible
algorithm for determining $E[x,y]$ even if $N$ is small enough to allow an
exhaustive survey of the landscape. The values of $E[x,y]$ and $B(x)$ can,
however, be retrieved from the \emph{barrier tree} of the landscape.  The
algorithm for constructing these barrier trees is presented in
\cite{Flamm:00a,Flamm:02a} (see also Ref. \cite{Ferreira:00a} for a
detailed account of the algorithm in the spin-glass context).  It is
implemented in the \texttt{barriers} program \footnote{The source code is
available at
\texttt{http://www.tbi.univie.ac.at/$\sim$ivo/RNA/Barriers/}.}, which
constructs the tree from a sorted list of energy values of all spin
configurations in the landscape. In a barrier tree, the leaves of the tree
represent the local minima, and the internal nodes represent the saddles,
with the barrier sizes given by the length of the branches connecting the
local minima to their corresponding saddles. Fig. \ref{fig:trr_diff}
illustrates typical barrier trees for the problems considered in this
paper.

\section{Depth and Difficulty}\label{sec:depth} 

The depth $D$ and difficulty $\psi$ of a landscape are measures of the
landscape structure that are directly related to the performance of
simulated annealing \cite{Hajek:88,Catoni:92,Kern:93,Ryan:95,Catoni:99}.
The {\em depth} of a landscape is defined as the maximum barrier height, $D
= \max B(x)$, where the maximum is taken over non-global minima only.  In
particular, it can be shown that simulated annealing converges almost
surely to a ground state if and only if the cooling schedule $T_k$
satisfies $\sum_{k\ge0}\exp(-D/T_k)=\infty$ \cite{Hajek:88}.  The
\emph{difficulty} of the landscape is a dimensionless quantity defined as
\begin{equation}
  \psi = \max \bigg\{ \frac{B(x)}{E(x)-E(\min)} \bigg\} 
\label{eq:difficulty}
\end{equation}
where $E(\min)$ is the global energy minimum and, as before, the maximum is
taken over local minima only.  The difficulty $\psi$ is directly related to
the optimal speed of convergence of simulated annealing. It is more
convenient to work with the scaled quantity
\begin{equation}
 \lambda = \log_2 \left ( \psi/2^N \right) = \log_2\psi - N 
\label{eq:lambda}
\end{equation}
instead.

\begin{figure}
\begin{center}
\begin{tabular}{cc}
\includegraphics[width=0.24\textwidth,clip=]{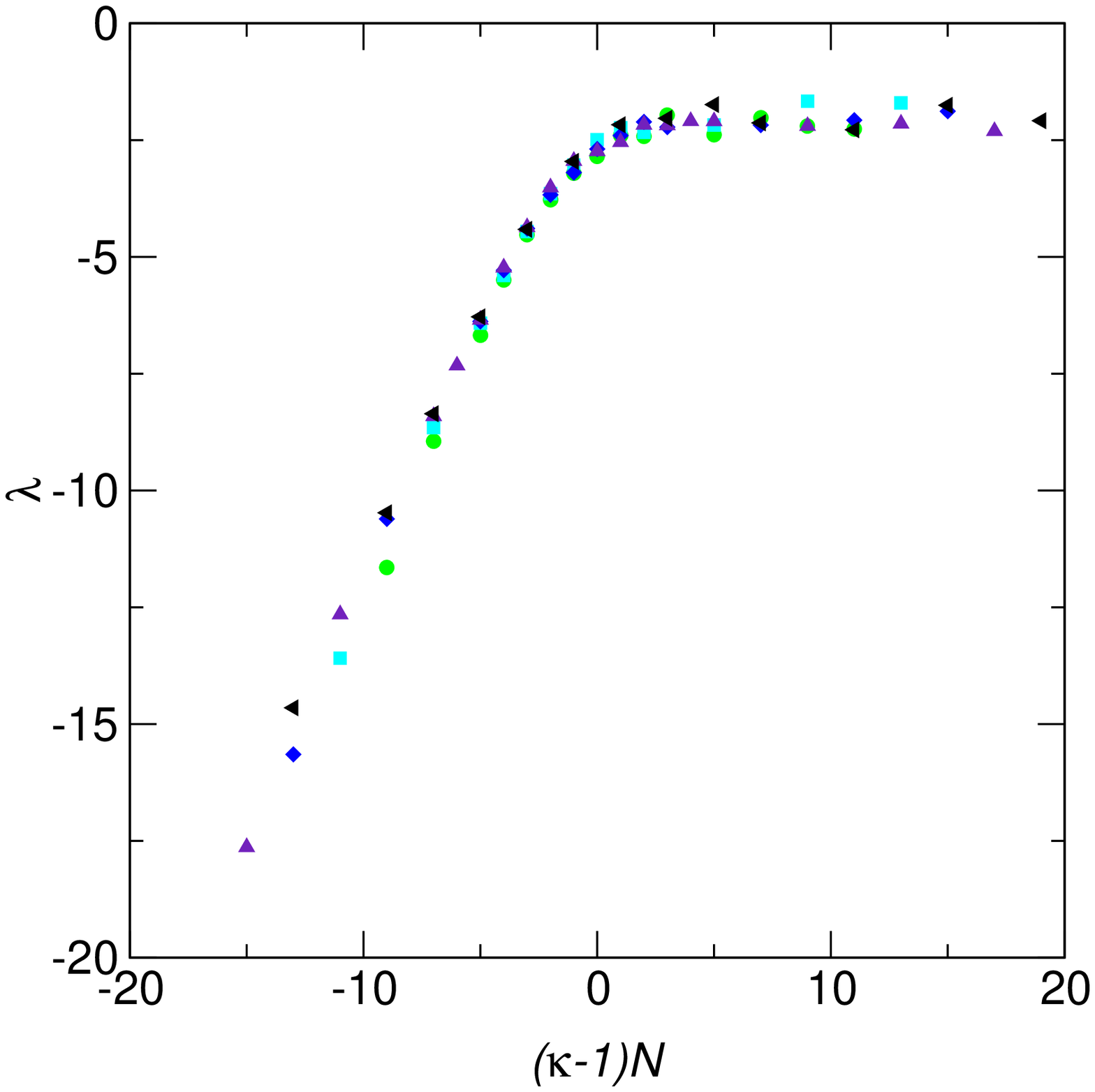} & 
\includegraphics[width=0.24\textwidth,clip=]{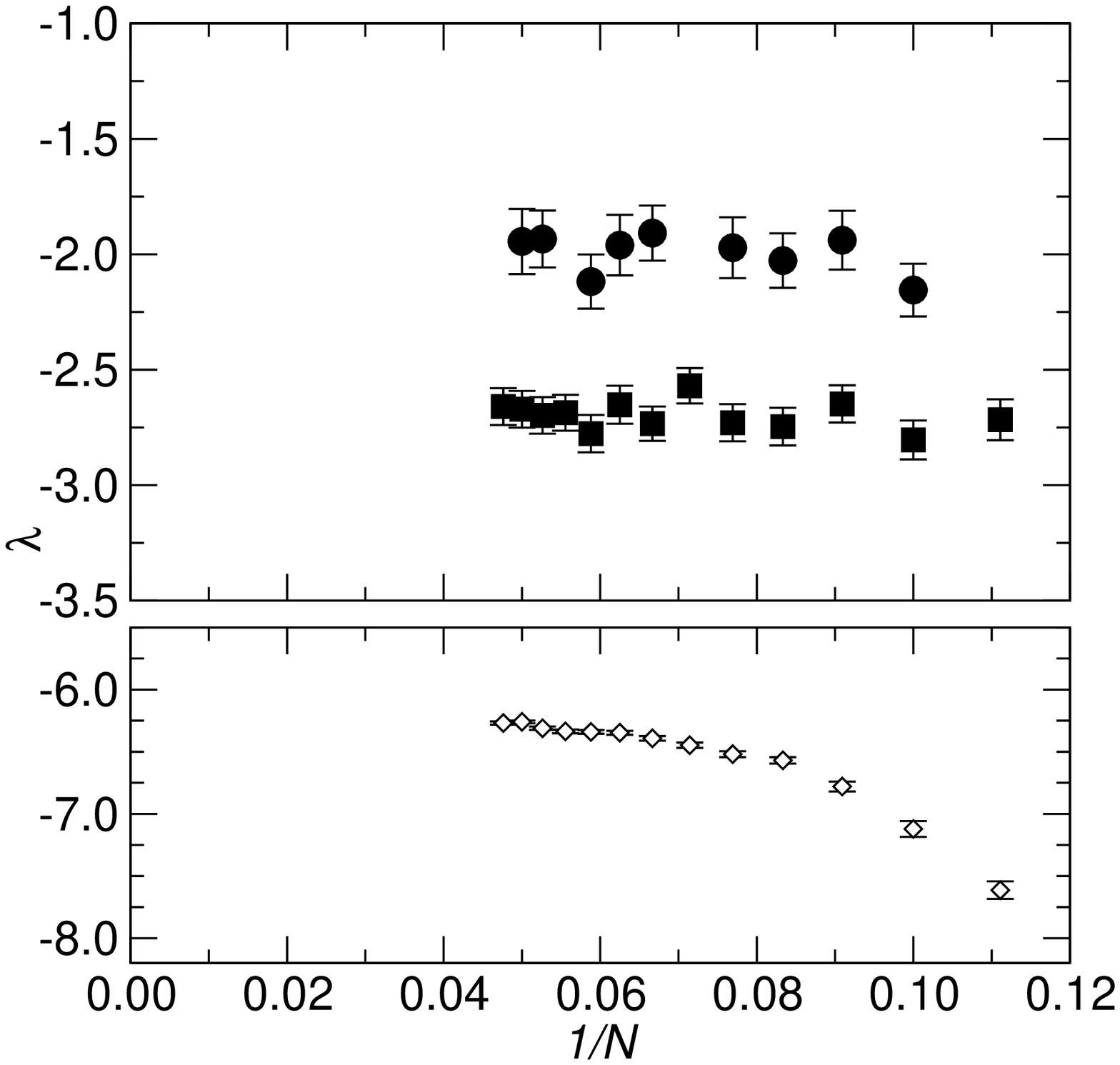} \\
{\small(a)} & {\small(b)}
\end{tabular}
\end{center}
\caption{(a) Data collapse of the rescaled average logarithmic difficulty
$\langle \lambda \rangle$ as function of the rescaled accuracy $(\kappa
-1)N$ for problem sizes $N=12,14,16,18$ and $20$. (b) Detailed data for
$(\kappa -1)N=5$ ($\bullet$), $0$ ($\blacksquare$), and $-5$ ($\diamondsuit$)
confirm the existence of the scaling function for large $N$.}
\label{fig:diff}
\end{figure}

We turn now to the evaluation of the effects on the depth and difficulty
measures of a change in the accuracy of the $a_i$s for a fixed problem
size. We find that, similarly to many other tree measures discussed in the
next section, the depth measure $D$ is independent of the accuracy of the
$a_i$s. The effects on the difficulty measure, on the other hand, are
striking. Explicitly, in Fig. \ref{fig:diff}, where each symbol represents
the result of the average over $100$ landscapes, we show that there is a
scaling relation between the average difficulty and the accuracy of number
representation: $\langle \lambda \rangle$ converges to a unique function of
$(\kappa - 1)N$ for large $N$.  This scaling function increases linearly
for $\kappa<1$ and approaches a constant value of about $-2.0$ for
$\kappa>1$. Hence $\langle \lambda \rangle$ viewed as a function of $\kappa
- 1$ exhibits a singularity at $\kappa = 1$ since it increases linearly
with a slope proportional to $N$ as long as $\kappa<1$ and tends towards a
constant value for $\kappa>1$. Therefore the rescaled difficulty reflects
the phase transition reported in previous analyses of the NPP, which have
focused on the singular behavior of the probability of a perfect partition
\cite{Gent:98,Mertens:98}.

A simple annealed-like argument to explain the behavior of the difficulty
depicted in Fig. \ref{fig:diff} goes as follows. In the easy regime we have
more partitions (spin configurations) than different combinations of
numbers, thus the global optimum will probably be a perfect partition,
i.e., $E_{\min} =0$ or $1$, while the lowest metastable state will have
$E=2$ or $3$. The height of the barrier separating them, however, is
essentially the energy of a random configuration, i.e.,
$B(x)=\mathcal{O}(l)$, and the maximum barrier height will be among the
largest numbers in the system, i.e., $\log_2\psi \approx \log_2 l$. Hence
subtracting $N$ from both sides yields $\lambda \approx (\kappa - 1) N$ in
the easy regime.  In the hard regime we can only hope to cancel the leading
$N$ bits in the optimal configuration, thus we expect a ground-state energy
$ E_{\min}\approx l/ 2^{N}$, while the maximum barrier height is again
$\mathcal{O}(l)$ yielding $\log_2\psi\approx N$, which is then independent
of $\kappa$.  Of course, these crude estimates miss polynomial corrections
such as the factor $N^{1/2}$ that appears in the rigorous computation of
the ground-state energy \cite{Karmarkar:86}. Note, however, that we are
considering $\log_2 E_{\min}$, i.e., we ignore only logarithmic corrections
which we would expect to arise for more careful estimates of $\log_2 B(x)$
as well.  As the values of $\ln\psi$ vary significantly among different
landscapes with the same values of $l$ and $N$ it is not possible to obtain
sufficiently accurate estimates of $\langle\lambda\rangle$ that would
reveal such corrections unambiguously.

\section{Measures of barrier tree shapes}\label{sec:shapes}

Since a barrier tree embodies all the relevant quantitative information
about the multi-valley structure of an energy landscape, it seems natural
to ask if there is any trace of the easy/hard phase transition in the shape
of the barrier trees of the NPP landscapes.

\begin{figure}
\begin{center}
\includegraphics[width=0.48\textwidth,clip=]{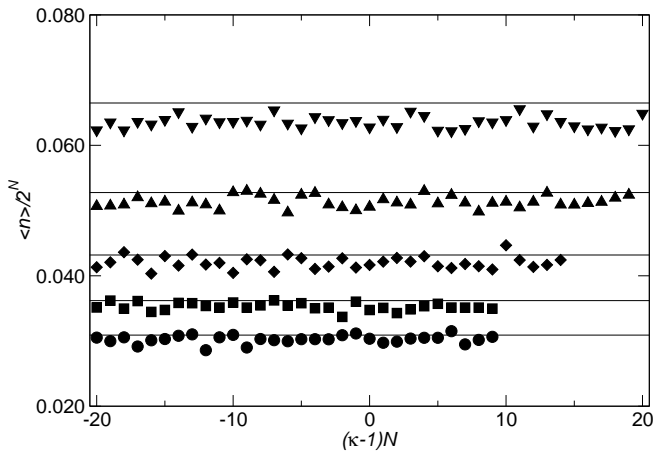}
\end{center}
\caption{Average fraction of local minima $\langle n \rangle/2^N$ as a
function of the rescaled accuracy $(\kappa -1)N$ for (top to bottom)
$N=12,14,16,18$ and $20$. Each symbol is the result of an average over 200
landscapes and the horizontal lines are the theoretical predictions for the
limit $l \to \infty$.}
\label{fig:locminA} 
\end{figure}

Before introducing the standard measures of tree shapes we will consider
the effect of the accuracy $l$ on probably the main characteristic of a
tree, namely, its number of leaves or minima $n$. In fact, it is tempting
to associate the difficulty of a problem with the number of local minima
(traps) in its energy landscape representation. Surprisingly, however, we
find that the average fraction of local minima does not change with $l$ as
soon as $l > N$, i.e., as soon as it becomes unlikely that adjacent
configurations have the same energy, see Fig.~\ref{fig:locminA}. For very
small values of $l$ the number of local optima depends strongly on how
degenerate neighbors are treated (data not shown).  The important point
here is that the fraction of minima stays constant across the easy/hard
transition and, in particular, it is given by the formula
\begin{equation}\label{eq:npp_min}
   \langle n \rangle/2^N = \sqrt{24/\pi}~N^{-3/2} ,
\end{equation}
which was obtained in the limit of infinite accuracy \cite{Ferreira:98}.

Now we consider five measures of tree shape that were originally used to
study phylogenetic trees (see, e.g., Refs.  \cite{Shao:90,Kirkpatrick:93})
for the barrier trees resulting from NPP landscapes. These measures provide
statistical information about the shape of the barrier tree, mainly its
symmetry or balance, and ignore the branch lengths, i.e., the height of the
barriers between minima, which were the object of the depth and difficult
measures.  Recently, we showed that these measures are capable of
distinguishing between $p$-spin models with different values of $p$
\cite{Hordijk:02}.

Let $d(i,j)$ be the graph-theoretical distance between two nodes of the
tree, i.e., the number of edges along the path that connects
them. Furthermore, we denote the root of the tree by $\varnothing$. The
\emph{height} of a leaf $k$ is $h_k=d(\varnothing,k)$. Equivalently, $h_k$
is the number of internal nodes between leaf $k$ and the root $\varnothing$
(inclusive). For each interior node $i$ we have two subtrees with $r_i$ and
$s_i$ leaves, respectively. We assume $r_i\geq s_i$. The
\emph{subtree-height} of an interior node $i$ is $m_i=\max_{k \in T_i}
d(i,k)$ where the maximum is taken over all leaves $k$ in the subtree $T_i$
below $i$, i.e., the subtree of which $i$ is the root.

With this notation we may define the following five characteristic values
for the shape of a binary rooted tree:
\begin{enumerate}
\item $H = \frac{1}{n} \sum_{k=1}^n h_k$ is the \emph{average height} of a
      leaf in the tree.
\item $\sigma_H = \sqrt{\frac{1}{n} \sum_{k=1}^n \left(h_k-H\right)^2}$,
      is the standard deviation of the leaf height.
\item $C = \frac{2}{n(n-3)+2} \sum_{i=1}^{n-1} \left(r_i-s_i\right)$ is a
      measure for the \emph{imbalance} of trees with $n > 2$.   
\item $B_1 = \sum_{i\ne\varnothing} 1/m_i$ is the average inverse
      subtree height, where the sum is taken over all $n-2$ internal nodes
      $i$ excluding the root $\varnothing$.
\item $B_2 = \sum_{k=1}^n 2^{-h_k}h_k$ is an
      alternatively weighted average leaf-height.
\end{enumerate}

The physical meaning of $H$ and $\sigma_H$ is clear. We mention only that
for random trees (e.g., trees produced by the neutral genealogical process
\cite{Felsenstein:02}) the expected value of $H$ increases as $\ln n$. In
addition, $\sigma_H = 0$ for a completely symmetric tree.  The imbalance
measure $C$ assigns a weight proportional to the the number of leaves to
each one of the two subtrees branching out an internal node.  These weight
differences are then averaged and normalized over all internal nodes of the
tree. The value of $C$ increases from 0 for a completely symmetric tree to
1 for a completely asymmetric tree.  The statistic $B_1$ looks at the
longest possible path $m_i$ from each internal node $i$ to any of the
leaves in its subtree.  The statistic $B_2$ is based on an index of
information content. For highly asymmetric trees, such as those produced by
$p$-spin landscapes \cite{Hordijk:02}, it will quickly converge to the
value $B_2 =2$.  Both $B_1$ and $B_2$ have smaller values for increasingly
asymmetric trees. 

\begin{figure}
\begin{center}
\includegraphics[width=0.48\textwidth,clip=]{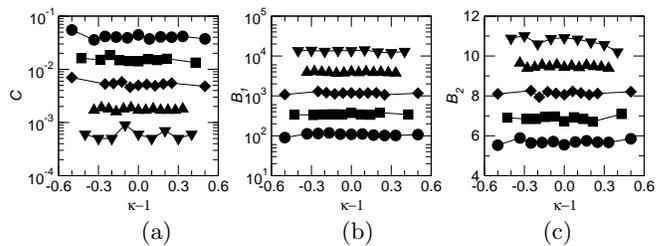}
\par
\qquad\parbox{0.32\textwidth}{(a)\hfill(b)\hfill(c)}
\end{center}
\caption{None of the measures of the tree-shape shown here reflect the
easy/hard transition of the NPP (a) imblance $C$, (b) inverse subtree
height $B_1$, (c) weighted leaf-height $B_2$. The data for $N=12
(\blacktriangledown), 14 (\blacktriangle), 16 (\blacklozenge), 18
(\blacksquare)$, and $20 (\bullet)$ are averages over 100 landscapes.}
\label{fig:treeshape}
\end{figure}

Somewhat surprisingly, neither of these measures
exhibits a non-trivial dependence on $\kappa$, as shown in
Fig.~\ref{fig:treeshape} for the measures $C$, $B_1$ and $B_2$. The other
quantities $H$ and $\sigma_H$ behave analogously but have a larger scatter
for large $N$.  Therefore it is impossible to tell the easy from the hard
regime by visual inspection of the barrier tree or by simply measuring
branch lengths, as done in the depth measure. The only effective measure
involves a nontrivial balance between branch lengths and leaf energies.

\section{Truncated  random energy model}\label{sec:TREM}

A very interesting though poorly explored finding concerns the equivalence
between the NPP in the limit of infinite precision, where the $a_i$ become
real random variables distributed uniformly in the unit interval, and the
symmetrized truncated random energy model (REM) or random cost problem
\cite{Mertens:00}. In particular, for large problem sizes the energies of
two or more distinct configurations of the NPP become statistically
independent and the $M = 2^{N-1}$ distinct energies values for
unconstrained partitions are distributed according to
\begin{equation}\label{eq:trem}
P(E) =
\frac{2}{\sqrt{2 \pi \mu_\infty N}}\exp\left(-\frac{E^2}{2 \mu_\infty N}\right)
  \Theta(E)
\end{equation}
where $\mu_\infty = \langle a ^2 \rangle = 1/3 $ is the second moment of
the $a_j$ in the corresponding number partitioning problem.  Hence the bold
claim that the NPP is essentially equivalent to a truncated REM
\cite{Mertens:00}. The main application of Eq. (\ref{eq:trem}) is the
derivation of the probability density of the minimum energy $E_{\min}$
using trite arguments of extreme statistics \cite{Mertens:00},
\begin{equation}\label{eq:rho}
\rho \left( E_{\min} \right ) = M P\left ( 0 \right )
  \exp\left[ - M P\left ( 0 \right ) E_{\min} \right] ,
\end{equation}
from which the expected minimum energy follows trivially,
\begin{equation}
\langle E_{\min} \rangle = \sqrt{2 \pi /3} ~N^{-1/2} ~2^{-N}
\end{equation}
in agreement with the known numerical \cite{Ferreira:98} and analytical
\cite{Mertens:98} results for the NPP.

However, in order for the equivalence at the level of the energy
distribution between the NPP and the truncated REM to have any use in
guiding the design of search heuristics for the NPP, it is important that
other features of the two problems, such as their multi-valley structures,
are similar too. In fact, a glance at Fig. \ref{fig:trr_diff} is already
sufficient to reveal the deep structural difference between the barrier
trees of these problems, and the remainder of this section is aimed at
quantifying these differences.

The expected number of minima can be easily calculated for any random
energy model with finite probability density over the reals
\cite{Derrida:81,Gross:84}.  The argument goes as follows. Fix an arbitrary
spin configuration $\sigma = (\sigma_1, \sigma_2, \ldots, \sigma_N)$ and
consider all its N neighbors. Since we assign to each spin configuration a
random energy value drawn from the continuous distribution (\ref{eq:trem})
we conclude that (i) all these $N+1$ energies are distinct with probability
1; and (ii) $E(\sigma)$ is the smallest of the N+1 numbers with a
probability of $1/(N+1)$. Hence $\sigma$ is a local minimum with
probability $1/(N+1)$ and so the fraction of local minima is
\begin{equation}
\langle n \rangle/2^N = 1/\left(N + 1 \right)
\end{equation}
which, for large $N$, is larger than the result for the NPP [see
Eq.(\ref{eq:npp_min})] by a factor of order of $N^{1/2}$.

\begin{figure}
\begin{center}
\includegraphics[width=0.50\textwidth,clip=]{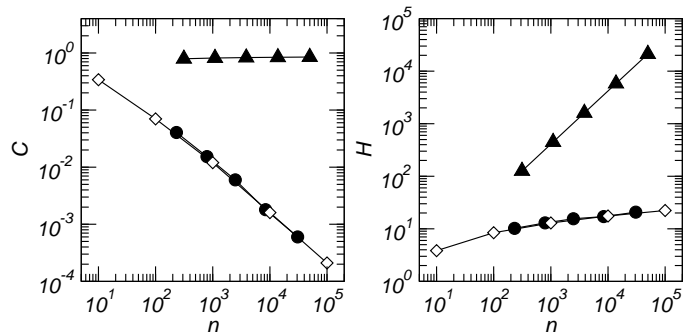}
\end{center}
\caption{Tree balance measures $C$ and $H$ as functions of the tree size
$n$ for the symmetrized truncated REM ($\blacktriangle$), the NPP
($\bullet$) and random trees ($\diamondsuit$). The solid lines are
numerical fittings: $H = 0.4 n$ and $ C = 0.6 + 0.1 \ln \ln \left( n
\right)$ for the REM, and $H = -1 + 2 \ln \left( n \right)$ and $ C = 5/n$
for the NPP and random trees.  }
\label{fig:new}
\end{figure}

In Fig.~\ref{fig:new} we present the tree size ($n$) dependence of the
barrier tree measures $C$ and $H$ for the NPP and the symmetrized truncated
REM.  A useful standard here, also shown in this figure, is the random
trees, generated as follows. First, create $n$ nodes (the leaves) and put
them in a set $A$. Next, remove two random nodes $x$ and $y$ from $A$,
create a new node $z$ and make $x$ and $y$ its two children, and put $z$ in
the set $A$. Repeat this procedure until there is only one node left in
$A$, which will be the root of the tree. Random trees are important from
the biological viewpoint because they arise from the neutral genealogical
process \cite{Felsenstein:02}.  The symmetrized truncated REM presents the
same scaling on $n$ as the $p$-spin models, namely, $ H \sim n$ and $ C
\sim \ln \ln \left (n \right )$ within the range of $n$ considered
\cite{Hordijk:02}. Of course, since $C \in \left [ 0,1 \right ]$ this
scaling cannot be valid for $n \to \infty$, but the double logarithmic
dependence guarantees its validity for a very large range of tree
sizes. Actually, as far as the five statistics introduced in
Sec.~\ref{sec:shapes} are concerned, there is no significative differences
between the symmetrized truncated REM and the standard REM, both models
producing then extremely unbalanced trees. These measures, however, differ
dramatically between the NPP and all previous spin-glass models analysed
\cite{Hordijk:02}. Surprisingly, the NPP barrier trees are practically
indistinguishable from random trees and, in particular, the tree measures
obey the same scaling relation with the tree size, $H \sim \ln \left ( n
\right )$ and $C \sim 1/n$. Hence these trees, in stark contrast with the
$p$-spin model and REM barrier trees, become more and more balanced as $n$
increases.

Since the equivalence between NPP and the truncated REM discussed above was
proved only in the case of infinite precision numbers (i.e., $\kappa \to
\infty$) it provides no clues about the easy/hard phase transition that
takes place at $\kappa = 1$.  Nevertheless, it is natural to ask if the
integer counterpart of the truncated REM, obtained by considering the
integer-valued energies $E'=\lfloor E \rfloor$ where $E$ is drawn from the
distribution (\ref{eq:trem}) with $\mu_\infty$ replaced by $\mu_l = l
(l+1)(2l+1)/6$, exhibits a phase transition. This can easily be answered by
calculating the probability that a partition is perfect (i.e., $E_{\min}' =
0$ or $1$) which is given by
\begin{equation}
P_{\mbox{\small{perf}}}  =   \int_0^2 dE_{\min} ~\rho
\left (E_{\min}  \right ) = 1 - \exp \left ( - 2 \xi \right )
\end{equation}
where $\xi = 2^N/\left ( 2 \pi \mu_l N \right)^{1/2}$. Although one clearly
recovers the easy ($P_{\mbox{\small{perf}}} \approx 1$) and the hard
($P_{\mbox{\small{perf}}} \approx 0$) regimes depending on whether the
ratio $(\log_2 l)/N$ is very small or very large, respectively, there is
definitely no phase transition separating them.

\section{Conclusion}

Phase transitions in physical systems are characterized by the appearance
of singularities in some observables, known as the order parameters of the
system, such as, e.g., the gas density in the boiling transition. In the
case of mean-field spin-glass models the order parameter directly reflects
the hierarchical organization of pure states in a complex multi-valley
structure \cite{Mezard:87}. Therefore one expects that some features of
that structure must undergo abrupt changes when the critical point is
approached. Unfortunately, the vast majority of the phase transitions take
place at finite temperature, while a direct study of the landscape
properties of spin-glass models based on natural quantities, such as saddle
points and minima, is feasible at zero temperature only. In that sense, the
easy/hard phase transitions in optimization problems \cite{Hogg:96} in
general, and in the number partitioning problem (NPP)
\cite{Gent:98,Mertens:98} considered here, provide a unique chance to study
how the onset of the phase transition affects the organization of the
metastable states of a disordered spin system.

Somewhat surprisingly, we find that almost all features of the landscape,
which we have properly mapped into a tree structure through the {\tt
barriers} program, are insensitive to the onset of the easy/hard phase
transition that takes place when the number of bits needed to specify a
number $a_i$ equals the problem size, i.e., $\kappa = (\log_2 l)/N =
1$. Interestingly, only one of the measures studied, premonitorily termed
{\it difficulty} in the mathematical literature of simulated annealing
\cite{Catoni:92}, exhibits a singular behavior at the critical point. As a
result, the quality of the optima found by simulated annealing will
probably depend strongly on whether the control parameters set the instance
in the easy ($\kappa <1$) or hard ($\kappa >1$) regime.

An important by-product of our study of the NPP landscape is the finding
that the resulting barrier trees are very similar to random trees, and so
they become completely balanced (symmetric) in the limit of large system
sizes $N$ or, equivalently, large tree sizes $n$. These trees contrast
drastically with the barriers trees resulting from the $p$ spin-glass, the
random energy or the symmetrized truncated random energy landscapes, which
become completely unbalanced in that limit. In this context, we note that
although there is an equivalence at the level of the energy distribution
between the NPP and the truncated random energy model, the statistical
properties of their energy landscapes are very different, and probably so
are the performances of local search heuristics in finding near-global
solutions to these problems. Actually, the similarity of the NPP barrier
trees with random trees may be part of the explanation for the failure of
local search techniques to produce good solutions to this optimization
problem.

\vspace{4mm}

\subsection*{Acknowledgments}
{\small Thanks to Christoph Flamm at the University of Vienna, Austria, for
his help with the \texttt{barriers} program. This research was supported by
Funda\c{c}\~ao de Amparo \`a Pesquisa do Estado de S\~ao Paulo (FAPESP),
project 99/09644-9. The work of J.F.F. is supported in part by CNPq and WH
is supported by FAPESP. P.F.S. gratefully acknowledges the hospitality of
the Instituto de F{\'\i}sica de S{\~a}o Carlos in Nov.\ 2002.  }

\end{document}